\documentclass[floats,aps,prd,nofootinbib,superscriptaddress,groupedaddress,showkeys]{revtex4}  
\usepackage{graphicx}  
\usepackage{dcolumn}   
\usepackage{bm}        
\usepackage{amssymb}   
\usepackage{amsmath}
\usepackage{xcolor}
\usepackage{bbold}
\usepackage{amsthm}
\newtheorem{theorem}{Theorem}
\newcommand{\be}{\begin{equation}}\newcommand{\ee}{\end{equation}}
\newcommand{\bea}{\begin{eqnarray}}\newcommand{\eea}{\end{eqnarray}}
\newcommand{\brr}{\begin{array}}\newcommand{\err}{\end{array}}
\newcommand{\bit}{\begin{itemize}}\newcommand{\eit}{\end{itemize}}
\newcommand{\ben}{\begin{enumerate}}\newcommand{\een}{\end{enumerate}}

\newcommand{\bbm}{\begin{bmatrix}}\newcommand{\ebm}{\end{bmatrix}}
\newcommand{\ba}{\begin{array}}
\newcommand{\ea}{\end{array}}

\newtheorem{mydef}{Definition}
\newtheorem{Lemma}{Lemma}
\newcommand{\bd}{\begin{mydef}} \newcommand{\ed}{\end{mydef}}
\newcommand{\bthe}{\begin{theorem}} \newcommand{\ethe}{\end{theorem}}
\newcommand{\ble}{\begin{Lemma}} \newcommand{\ele}{\end{Lemma}}

\newcommand{\dr}{\mathrm{d}}

\definecolor{darkred}{rgb}{.8,0,0}

\definecolor{darkblue}{rgb}{0,0,.7}

\def\lan{\langle}
\def\lf{\left}

\def\non{\nonumber}\def\ran{\rangle}

\def\ri{\right}
\def\al{\alpha}\def\bt{\beta}
\def\de{\delta}\def\De{\Delta}

\def\la{\lambda}\def\si{\sigma}\def\Si{\Sigma}

\def\1{{_{1}}}\def\2{{_{2}}}

\def\noHe0{:\;\!\!\;\!\!:H_e(0):\;\!\!\;\!\!:}
\def\noHm0{:\;\!\!\;\!\!:H_\mu(0):\;\!\!\;\!\!:}

\def\lan{\langle}
\def\lf{\left}

\def\non{\nonumber}
\def\ran{\rangle}

\def\ri{\right}

\def\al{\alpha}\def\bt{\beta}
\def\de{\delta}\def\De{\Delta}

\def\la{\lambda}
\def\si{\sigma}\def\Si{\Sigma}

\def\1{{_{1}}}\def\2{{_{2}}}

\begin{document}

\title{Time-energy uncertainty relation for neutrino oscillations in curved spacetime}
\author{M.~Blasone}
\email{blasone@sa.infn.it}
\affiliation{Dipartimento di Fisica, Universit\`a di Salerno, Via Giovanni Paolo II, 132 84084 Fisciano, Italy}
\affiliation{INFN Sezione di Napoli, Gruppo collegato di Salerno, Italy}
\author{G.~Lambiase}
\email{lambiase@sa.infn.it}
\affiliation{Dipartimento di Fisica, Universit\`a di Salerno, Via Giovanni Paolo II, 132 84084 Fisciano, Italy}
\affiliation{INFN Sezione di Napoli, Gruppo collegato di Salerno, Italy}
\author{G.~G.~Luciano}
\email{gluciano@sa.infn.it}

\affiliation{INFN Sezione di Napoli, Gruppo collegato di Salerno, Italy}
\author{L.~Petruzziello}
\email{lpetruzziello@na.infn.it}

\affiliation{Dipartimento di Fisica, Universit\`a di Salerno, Via Giovanni Paolo II, 132 84084 Fisciano, Italy}
\affiliation{INFN Sezione di Napoli, Gruppo collegato di Salerno, Italy}
\author{L.~Smaldone}
\email{lsmaldone@sa.infn.it}
\affiliation{INFN Sezione di Napoli, Gruppo collegato di Salerno, Italy}
%
\vspace{3mm}

\begin{abstract}
We derive the
Mandelstam-Tamm time-energy uncertainty relation
for neutrino oscillations in a generic stationary curved spacetime.
In particular,  by resorting to Stodolsky covariant formula of the quantum mechanical 
phase, we estimate gravity effects on the neutrino energy 
uncertainty. Deviations from the standard Minkowski result 
are explicitly evaluated in Schwarzschild, 
Lense-Thirring and Rindler (uniformly accelerated) geometries. 
Finally, we discuss how spacetime could affect the characteristic neutrino oscillation length in connection with the recent view
of flavor neutrinos as unstable particles.
\end{abstract}


\maketitle

\section{Introduction}
Neutrino mixing and oscillations are among
the most active research areas within the 
framework of particle physics. Since Pontecorvo's 
original treatment in quantum mechanics (QM)~\cite{Pontecorvo}, 
the theoretical bases of these phenomena 
have been extensively analyzed~\cite{partref}, and a 
quantum field theoretical (QFT) formalism has been 
developed~\cite{Blas95, qftmixing}, providing continuous 
insights toward understanding of novel 
effects~\cite{Dolgov:2002wy,Blasone:2007vw,Akhmedov:2010ua}. 
In spite of this, such puzzling questions 
as the dynamical origin of the non-vanishing neutrino masses and mixings in the Standard Model
still remain open~\cite{Blasone:2018hah}.

Starting from the results of Ref.~\cite{Bhat}
in the context of unstable particle physics, 
in Refs.~\cite{Bil} it was argued that
the Mandelstam-Tamm time-energy 
uncertainty relation (TEUR)~\cite{ManTam}
reduces to the well-known condition for neutrino oscillations
described by Pontecorvo states~\cite{Kayser}.
Such an outcome was recently 
revisited in Ref.~\cite{smald}, where it was found
that TEUR can be rephrased in terms of a \emph{flavor-energy} uncertainty principle 
as a consequence of the non-commutativity between
the Hamiltonian and the lepton charge operators~\cite{Blas95, qftmixing}.
In that case, flavor-energy uncertainty relations imply that neutrino energy cannot be sharply determined, and the study of its distribution reveals to be important from both theoretical and phenomenological viewpoints.  
Furthermore, following the quantum field theoretical approach 
of Refs.~\cite{Blas95, qftmixing}, 
the aforementioned oscillation condition 
was interpreted as a fundamental bound 
on neutrino energy precision: as for unstable particles, 
only energy distributions are meaningful for flavor neutrinos 
and the width of the distribution is related to the oscillation length. 

The above analysis has been only performed 
in flat spacetime. The first attempt to accommodate gravity effects
into the standard picture of neutrino oscillations
was made in Ref.~\cite{Stodo}.
Further investigation was later carried out
in Refs.~\cite{Aluw, Cardall,Forne}, showing that gravity-induced
corrections are related to 
the energy redshift.
Recently, similar results were derived
in accelerated frames in both QM~\cite{Blasone:2018iih,acc}
and QFT~\cite{Blasone:2017nbf}.
The question thus arises as to how the
framework of neutrino oscillations and, in particular, 
the condition on the energy uncertainty stemming from TEUR 
is modified in curved backgrounds.

Based on the outlined scenario, 
in the present paper we derive the 
Mandelstam-Tamm version of TEUR
for neutrino oscillations in a generic 
stationary curved spacetime.  To this aim, we
employ the quantum mechanical covariant formalism for oscillations
developed in Ref.~\cite{Cardall}. 
Corrections to the lower bounds on the neutrino 
energy uncertainty are explicitly evaluated for the cases of
Schwarzschild, Lense-Thirring and Rindler (i.e. uniformly accelerated) 
backgrounds. In particular, we find that the characteristic oscillation length gets non-trivially modified by gaining additional terms which depend on the specific features of the considered geometry. In light of neutrino interpretation as unstable particle, this amounts to state that neutrino lifetime is not an intrinsic quantity. Note that such a result is not entirely uncommon in literature; in Ref.~\cite{Mueller}, indeed, it was argued that the decay properties of particles are not fundamental, since they depend on the reference frame.

The paper is organized as follows: in Section~\ref{man} 
we prove that Mandelstam-Tamm TEUR can be derived 
in a generic stationary curved spacetime. 
In Section~\ref{teurno}, after a brief review of the standard
treatment of neutrino oscillations within Minkowski framework, 
we discuss the covariant approach of Ref.~\cite{Cardall}.
We also show how to recast TEUR in a form suitable 
for generalization to curved backgrounds. 
Section~\ref{teurcs} is devoted to the calculation of the neutrino 
energy uncertainty in Schwarzschild, Lense-Thirring and Rindler 
spacetimes. In order to separate metric-induced corrections out, 
we adopt the weak gravity/acceleration approximation.
Finally, conclusions and discussion can be found in Section~\ref{conc}.

Throughout the work we shall assume natural units $\hbar=c=1$
and the conventional timelike signature for Minkowski metric,
\be
\label{minkmet}
\eta_{\mu\nu}=\mathrm{diag}[1,-1,-1,-1] .
\ee
Furthermore, we shall denote local
(general curvilinear) coordinates by indices with (without) hat.
\section{Mandelstam-Tamm TEUR in stationary spacetime}
\label{man}
In this Section, we show how to generalize the Heisenberg
equation of motion
\be \label{he}
\frac{\dr}{\dr t}\hspace{0.2mm}O(x)=i\lf[H,O(x)\ri]
\ee
to the case of a generic stationary curved background. 
For this purpose, let us observe that, 
for stationary metrics, there exists a global timelike 
Killing vector field $K^\mu$, such 
that the quantity
\be
\label{K}
K \equiv  \int K_\mu\, T^{\mu \nu}\, \dr \Si_\nu\,
\ee
does not depend on the choice of the spacelike
hypersurface $\Sigma$ if $T^{\mu\nu}$ is the (conserved) stress-energy
tensor~\cite{dw}. Since $K^\mu$ is globally timelike, 
one can introduce a coordinate $t$ upon which
the metric does not depend and with respect
to which $K^{\mu}$ can be written as $K^{\mu}=(1,0,0,0)$.
In this case, Eq.~\eqref{K} can be recast as
\be
\label{H}
K \equiv  \int :T^{00}: \dr^3x=H\,,
\ee
where we have introduced the normal ordering.
Moreover, $K$ satisfies the relation~\cite{dw}
\be
\label{lie}
\lf[O(x),K\ri]=i\mathcal{L}_K O(x),
\ee
where $\mathcal{L}_K$ 
is the Lie derivative with respect to $K_\mu$.
This is exactly the Heisenberg equation~\eqref{he},
since $\mathcal{L}_K O(x) \equiv~\hspace{-1mm}\dr\hspace{0.2mm}O(x)/\dr t$ in the present case.

Using the Cauchy-Schwarz inequality, from Eq.~\eqref{he}
it follows that the~\emph{Mandelstam-Tamm} version of TEUR~\cite{ManTam} in a generic
stationary background
can be formulated as
\be \label{teunc}
\Delta E \, \Delta t \, \geq \frac{1}{2} \, ,
\ee
with 
\be
\Delta E \equiv \si_H,  \quad\,\,\, \Delta t \equiv \si_O\Big/\lf|\frac{\dr\langle O\rangle}{\dr t}\ri|,
\label{teunc1}
\ee
where $\sigma_A$ denotes the standard deviation of the generic operator $A$, i.e.
\be
\sigma_A=\sqrt{\langle A^2\rangle-\langle A\rangle^2}\,.
\ee
In the above expressions, we have used the shorthand notation
$\lan \ldots \ran \equiv \lan \psi| \ldots |\psi\ran$, 
where $|\psi\rangle$ denotes the state of 
the system, $O(t)$ represents the ``clock observable'' 
that regulates time changes through its 
dynamics and $\Delta t$ is the characteristic 
time interval that the mean value of $O$ takes 
to vary by a standard deviation $\sigma_O$~\cite{Schumacher}.
\section{TEUR for neutrino oscillations}\label{teurno}
Let us now derive TEUR for neutrino oscillations in a
generic stationary curved background.
As a first step, we review the standard 
analysis of flavor oscillations
in Minkowski spacetime. Then, we discuss  
the covariant formalism based on Stodolsky 
definition of quantum mechanical phase~\cite{Stodo}.
We restrict our analysis
to a simplified two-flavor model. However, the obtained result can be
straightforwardly generalized to a more rigorous three-flavor description.

\subsection{Neutrino oscillations in Minkowski spacetime}
\label{NoiMs}
In the standard treatment of neutrino oscillations, 
the flavor eigenstates $|\nu_\alpha\rangle$ are written
as a superposition of the mass eigenstates $|\nu_k\rangle$ 
according to~\cite{partref}
\be \label{eqn:genstate}
|\nu_\al(x)\ran =  \sum_{k=1,2} \, U_{{\alpha k}}\hspace{0.1mm}(\theta) \, |\nu_k(x)\ran \, , \quad \al=e,\mu \, , 
\ee
where $U_{{\alpha k}}\hspace{0.1mm}(\theta)$ is the generic element 
of the Pontecorvo mixing matrix,
\begin{equation}
\label{PMM}
U(\theta)=
\begin{pmatrix}
\cos\theta&\sin\theta\\
-\sin\theta&\cos\theta
\end{pmatrix}\,.
\end{equation}
The spacetime evolution of mass eigenstates is described by\footnote{In order to streamline the notation, we shall omit the spacetime dependence when there is no ambiguity.}
\begin{equation}
\label{defmas}
|\nu_k({x})\rangle=\exp[{-i\varphi_k(x)}]\hspace{0.2mm}|\nu_k\rangle,\quad (k=1,2)\,,
\end{equation}
where
\begin{equation}
\label{phase}
\varphi_k(x)\equiv E_k\hspace{0.2mm}t-\textbf{p}_k\cdot\textbf{x}
\end{equation}
is the quantum-mechanical phase of the $k^{th}$ state 
with mass $m_k$, energy $E_k$ and three-momentum $\textbf{p}_k$, respectively. 
Mass, energy and momentum are related by the usual
mass-shell condition as
\begin{equation}
E^2_k=m^2_k+{|\textbf{p}_k|}^2\,.
\end{equation}
The oscillation (survival) probability 
$\mathcal{P}_{\al\rightarrow \bt}$ ($\mathcal{P}_{\al\rightarrow \al}$) 
that a neutrino produced  as $|\nu_\alpha\rangle$ 
at a spacetime point $x_0$ is detected 
as $|\nu_\beta\rangle$ ($|\nu_\al\rangle$) at a point $x$ 
is, therefore, 
\bea\label{pontosc}
\mathcal{P}_{\al\rightarrow \bt}(x) & = & |\lan \nu_\bt(x)|\nu_\al(x_0)\ran|^2 \\[2mm]
&=& \lan \nu_\al(x_0)|\mathbb{P}_\bt(x)|\nu_\al(x_0)\ran \non \\[2mm] \non
& = & \sin^2 (2 \theta)\, \sin^2 \lf(\frac{\varphi_{12}}{2}\ri)\hspace{-0.7mm}, \,\,\,\,\, \alpha\neq\beta\,,\\[2mm]
\mathcal{P}_{\al\rightarrow \al}(x) \hspace{1mm}&=& 1- \mathcal{P}_{\al\rightarrow \bt}(x)\,,
\end{eqnarray}
where $\varphi_{12}\equiv \varphi_1-\varphi_2$ and $\varphi_k$ $(k=1,2)$ are the phases acquired by the mass eigenstates during their propagation, i.e.
\be
\varphi_k=E_k\,(t-t_0)-\textbf{p}_k\cdot(\textbf{x}-\textbf{x}_0)\,.
\ee
Furthermore, we have defined 
\be
\mathbb{P}_\bt(x) \equiv |\nu_\bt(x)\ran \lan \nu_\bt(x)| \,
\ee
the projector on the neutrino flavor state $|\nu_\bt(x)\ran$. 

Now, let us take $O(x)=\mathbb{P}_\al(x)$  
and $|\psi\ran=|\nu_\al(x_0)\ran$ in Eq.~\eqref{teunc}. 
By simple calculations, it is possible to show that~\cite{Akhmedov:2008zz}
\be \label{neutunqm}
2\hspace{0.3mm}\Delta E  \,\sqrt{\mathcal{P}_{\al\rightarrow \al}(x)\left[1-\mathcal{P}_{\al\rightarrow \al}(x)\right]}\geq\lf|\frac{\dr \mathcal{P}_{\al\rightarrow \al}(x)}{\dr t}\ri| \, .
\ee
By observing that the l.h.s. of Eq.~\eqref{neutunqm} reaches its maximum
when $\mathcal{P}_{\al\rightarrow \al}=1/2$, we then obtain
\be 
\Delta E\geq\lf|\frac{\dr \mathcal{P}_{\al\rightarrow \al}(x)}{\dr t}\ri| \, .
\ee
If we now integrate both sides over time  
and employ the triangular inequality, 
the above relation reads~\cite{smald}
\be
\label{triang}
\Delta E \, T  
\geq  \,\lf|\int^t_{t_0} \frac{\dr\mathcal{P}_{\al\rightarrow \al}(x)}{\dr t'}\, \dr t' \ri| =\mathcal{P}_{\al\rightarrow \beta}\hspace{0.5mm}(x)\,,
\ee
where $T \equiv \int_{t_0}^{t}dt'=t-t_0$
provides the neutrino time of flight.

In order to evaluate the oscillation probability 
$\mathcal{P}_{\al\rightarrow \beta}$, let us consider, for
simplicity, relativistic neutrinos propagating along a fixed direction 
(for example, the $x$-axis).
In this case, one can check that the phase Eq.~\eqref{phase} 
takes the form
\begin{equation}
\label{eqn:phasebis}
\varphi_{k}\simeq\frac{m^2_k}{2E
}\hspace{0.2mm}L_p\,,
\end{equation}
where we have assumed neutrino mass eigenstates to be
energy eigenstates with a common energy
$E
$,\footnote{For a more detailed discussion 
on whether neutrino mass eigenstates should be
taken as momentum eigenstates, energy eigenstates, or
neither, see, for example,  Ref.~\cite{gross}. These technicalities, however, 
do no affect our analysis.} and 
\be
\label{lpt}
L_p\equiv x-x_0\simeq T
\ee
is the proper distance travelled by neutrinos. 
Therefore, from Eqs.~\eqref{eqn:genstate} and \eqref{defmas} we have
\begin{equation}
\label{relapp}
|\nu_\alpha({x})\rangle=
\sum_{k=1,2}U_{{\alpha k}}\hspace{0.1mm}(\theta)\hspace{0.2mm}\exp\left({-i\hspace{0.3mm}\frac{m^2_k}{2E
}\hspace{0.3mm}L_p}\right)\hspace{-0.4mm}|\nu_k\rangle\,. 
\end{equation}
By virtue of this relation, the oscillation probability~\eqref{pontosc} 
becomes
\be
\label{oscpr}
\mathcal{P}_{\al\rightarrow \bt}(x) \ = \ \sin^2 (2 \theta)\sin^2 \lf(\frac{\pi\hspace{0.2mm}L_p}{L^{osc}}\ri),
\ee
where we have defined the characteristic \emph{oscillation length} $L^{osc}$ as
\be
\label{oscleng}
L^{osc}\equiv\frac{4 \pi E
}{\De m^2}, 
\ee
with $\De m^2 = m_2^2-m_1^2$. By inserting into Eq.~\eqref{triang}, this allows us to derive the following time-energy uncertainty condition for neutrino oscillations:
\be
\label{teurriformulation}
\Delta E \, T  
\geq  \sin^2 (2 \theta)\sin^2 \lf(\frac{\pi \hspace{0.2mm} L_p}{L^{osc}}\ri).
\ee
Since the above inequality holds true for any spacetime point, 
we can set $x$ in such a way to maximize the r.h.s., which implies 
$L_p= L^{osc}/2$. By means of Eq.~\eqref{lpt}, 
this yields
\be \label{minkteur}
\De E \geq \frac{2 \, \sin^2(2 \theta)}{L^{osc}} \, .
\ee
Note that, since $H$ is time-independent,
the same holds for $\Delta E$. Furthermore, for $L^{osc}\to \infty$,  
we obtain a vanishing lower bound for
$\Delta E$, as one could expect in the absence of flavor oscillations. The same is true for $\theta=0$: in that case, indeed, neutrino states with definite flavor do coincide with the states with definite mass, which are characterized by a well-defined energy.
From the point of view of Ref.~\cite{smald}, this means that neutrinos can be described by standard asymptotic states and not as unstable particles. Therefore, flavor and energy are commuting quantities and no fundamental bounds on energy measurements can be inferred.  

Inequalities of the form~\eqref{minkteur} are widely discussed in literature and are usually interpreted as conditions of neutrino oscillation~\cite{Kayser,Bil,smald}.

\subsection{Neutrino oscillations in curved spacetime} 
 \label{cs}
Let us now extend the above formalism
to curved backgrounds. To this aim,  it comes in handy 
to cast Eqs.~\eqref{eqn:genstate} and (\ref{relapp})
in a manifestly covariant form.  Following Ref.~\cite{Cardall}, 
we can then write
\be
\label{covfor}
|\nu_\alpha({\lambda})\rangle=\sum_{k=1,2} U_{{\alpha k}}\hspace{0.1mm}(\theta)\hspace{0.2mm}\exp(-i\Phi)|\nu_k\rangle,
\ee
where $\Phi$ is defined as
\be
\label{phasep}
\Phi=\int_{\lambda_0}^\lambda P_{\mu}\frac{\dr x^\mu_{null}}{\dr \lambda'}\hspace{0.2mm}\dr\lambda'\,.
\ee
Here, $P_\mu$ is the generator of spacetime
translation of neutrino mass eigenstates and $\dr x^\mu_{null}/d\lambda$ 
is the null tangent vector to the neutrino worldline parameterized by $\lambda$.

It is easy to show that Eq.~\eqref{covfor}
reduces to Eq.~\eqref{relapp} for neutrino propagation 
in flat spacetime. The advantage, however,
is that it allows us to generalize the study of
flavor oscillations to curved backgrounds.
Indeed, consider the covariant Dirac equation for a doublet $\nu$
of spinors of different masses~\cite{Weinberg:1995mt},
\begin{equation}
\label{eqn:covdireq}
\Big[i\gamma^{a}\hspace{0.2mm}e^{\mu}_{\hat{a}}\left(\partial_\mu+\Gamma_\mu\right)-M\Big]\nu=0\,,
\end{equation}
where $M=\mathrm{diag}[m_1,m_2]$ is the mass matrix in neutrino 
mass basis and $\gamma_\mu$ are the Dirac matrices satisfying the Clifford
algebra $\{\gamma^{\mu},\gamma^{\nu}\}= 2 g^{\mu\nu} \mathbb{1}\,.$
The general curvilinear and locally inertial coordinates 
are connected by the 
tetrads $e^{\mu}_{\hat{a}}$, as usual. 

The above equation can be simplified by
observing that the only non-vanishing contribution 
of the spin connection $\Gamma_\mu=\frac{1}{8}\left[\gamma^{\hat{b}},\gamma^{\hat{c}}\right]\hspace{-0.5mm}e^{\nu}_{\hat{b}} \hspace{0.2mm}\nabla_\mu e_{{\hat{c}}\hspace{0.2mm}\nu}$
is
\begin{equation}
\gamma^{\hat{a}}\hspace{0.1mm}e^{\mu}_{\hat{a}}\hspace{0.1mm}\Gamma_\mu=\gamma^{\hat{a}}\hspace{0.1mm}e^{\mu}_{\hat{a}}\left[i\hspace{0.1mm}A_{\mu}\left(-\,{g}^{-1/2}\hspace{0.2mm}\frac{\gamma^5}{2}\right)\right]\,,
\end{equation}
where $g\equiv|\mathrm{det}\,g_{\mu\nu}|$, $\gamma^5=i\gamma^0\gamma^1\gamma^2\gamma^3$
and
\begin{equation}
\label{eqn:A}
A^{\mu}=\frac{\sqrt{g}}{4}\hspace{0.3mm}\hspace{0.3mm}e^{\mu}_{\hat{a}}\hspace{0.3mm}\epsilon^{{\hat{a}}{\hat{b}}{\hat{c}}{\hat{d}}}\left(\partial_\sigma e_{{\hat{b}}\nu}-\partial_\nu e_{{\hat{b}}\sigma}\right)e^{\nu}_{\hat{c}} e^{\sigma}_{\hat{d}}\,,
\end{equation}
with $\epsilon^{{\hat{a}}{\hat{b}}{\hat{c}}{\hat{d}}}$ being the totally antisymmetric tensor
of component $\epsilon^{{\hat{0}}{\hat{1}}{\hat{2}}{\hat{3}}}=+1$. 

The momentum operator 
$P_\mu$ appearing in Eq.~\eqref{phasep} can be
derived from the generalized mass-shell relation, 
\begin{equation}
\label{eqn:genmascond}
\left(P^{\mu}+\frac{A^\mu\gamma^5}{2}\right)\left(P_{\mu}+\frac{A_{\mu}\gamma^5}{2}\right)=M^2.
\end{equation}
Neglecting terms of $\mathcal{O}(A^2)$ and $\mathcal{O}{(AM^2)}$, 
one gets that the quantity $P_{\mu}{\dr x^\mu_{null}}/{\dr \lambda}$
for relativistic neutrinos moving 
along generic null trajectories parameterized by $\lambda$ reads 
\begin{equation}
\label{eqn:pmudxmu}
P_\mu\hspace{0.3mm}\frac{\dr x^\mu_{null}}{\dr \lambda}=\left(\frac{M^2}{2}-\frac{\dr x^\mu_{null}}{\dr \lambda}A_{\mu}\gamma^5\,\right).
\end{equation}
In deriving Eq.~\eqref{eqn:pmudxmu}, we have required the three-momenta of $P^\mu$
and ${\dr x^\mu_{null}}/{\dr \lambda}$ to be parallel (i.e. $P^i\simeq {\dr x^i_{null}}/{\dr \lambda}$) and $P^0={\dr x^0_{null}}/{\dr \lambda}$~\cite{Cardall}.

By substituting the above expression in Eq.~(\ref{phasep}), the covariant phase
takes the form
\be\label{phasep2}
\Phi=\int_{\lambda_0}^\lambda \left(\frac{M^2}{2}-\frac{\dr x^\mu_{null}}{\dr \lambda'}A_{\mu}\gamma^5\,\right)\hspace{0.2mm}\dr\lambda'\,,
\ee
where $\dr\lambda$ can be written 
in terms of the differential proper distance at constant time
$\dr \ell$ by using the condition of null trajectory,
\begin{equation}
\label{dl}
\dr\lambda  =  \dr \ell{\left(\hspace{-1mm}-g_{ij}\frac{\dr x^i}{\dr \lambda}\frac{\dr x^j}{\dr \lambda}\right)}^{\hspace{-1mm}-\frac{1}{2}}\hspace{-2mm}\\[2mm] =  \dr \ell{\left[g_{00}\lf(\frac{\dr t}{\dr \lambda}\ri)^2\hspace{-1.5mm}+2\hspace{0.3mm}g_{0i}\hspace{0.3mm}\frac{\dr t}{\dr \lambda}\hspace{0.3mm}\frac{\dr x^i}{d\lambda}\right]}^{-\frac{1}{2}}\hspace{-3mm}.
\end{equation}
%
%
Exploiting the above formalism, we can now
recast TEUR  in a form suitable
for generalization to arbitrary stationary backgrounds. 
In this perspective, let us rephrase Eq.~\eqref{triang} in 
terms of the affine parameter $\lambda$,
\be
\Delta E \, T(\la)   \geq  \lf|\int^\la_{\la_0}\frac{\dr\mathcal{P}_{\al\rightarrow \al}(x(\la'))}{\dr \la'}\,\dr \la' \ri| = \mathcal{P}_{\al\rightarrow \bt}\hspace{0.3mm}(x(\lambda))  \, ,
\ee
where $T(\lambda)\equiv t(\lambda)-t(\lambda_0)$ and 
the integration has been performed over the neutrino null
worldline. By use of Eqs.~\eqref{pontosc}, this yields
\be\label{sp}
\Delta E \geq \frac{\mathrm{sin}^2\lf(2\theta\ri)}{T(\lambda)}\,\mathrm{sin}^2\bigg[\frac{{\varphi_{12}(x(\lambda))}}{2}\bigg], \quad {\varphi}_{12}\equiv{\varphi}_1-{\varphi}_2\,,
\ee
where now $\varphi_k$ ($k=1,2$) is more generally  
defined by the action of the operator
$\Phi$ in Eq.~\eqref{phasep2} on the $k^{th}$ neutrino state, i.e.
\be
\label{om12}
\Phi|\nu_k\rangle=\varphi_k|\nu_k\rangle.
\ee 
It should be emphasized that, in the above treatment,  
$\Delta E$ is the energy uncertainty 
for an observer at rest at infinity. This holds true since $E$ is the
eigenvalue of the Hamiltonian $H$ in 
Eq.~\eqref{he}, which is
the time component  
of the covariant momentum $P_\mu$ (see footnote 2). 
In order to recast Eq.~\eqref{sp} in terms of
the energy uncertainty $\Delta E_\ell$ for a 
local observer momentarily at rest
in the curved background 
(and also with
respect to the oscillation experiment), 
we must employ vierbein fields, so that
\be
\label{ene2}
E_\ell(x)\equiv P_{\hat{0}}(x)=e^{\nu}_{\hat{0}}(x)\, P_{\nu}\,.
\ee
Furthermore, the phase-shift $\varphi_{12}$
must be expressed in terms of the local energy 
and proper distance.

\section{TEUR in curved spacetime} \label{teurcs}
Let us apply Eq.~\eqref{sp} to 
estimate the energy uncertainty 
of neutrinos propagating in non-trivial stationary 
backgrounds. 
Specifically, we consider the cases of 
Schwarzschild, Lense-Thirring and Rindler 
metrics. In our calculations, we adopt the weak field
approximation, so that deviations of the energy uncertainty 
from the corresponding flat expression~\eqref{minkteur} can
be explicitly exhibited.
\subsection{Schwarzschild metric}
It is well known that, in isotropic coordinates, 
the (linearized) weak field Schwarzschild metric reads 
\be\label{schw}
\dr s^2=\lf(1+2\phi\ri)\dr t^2-\lf(1-2\phi\ri)\lf(\dr x^2+\dr y^2+\dr z^2\ri)\, ,
\ee
where the gravitational potential $\phi$ is defined as
\be\label{phi}
\phi(r)=-\frac{GM}{r}\equiv-\frac{GM}{\sqrt{x^2+y^2+z^2}},
\ee
with $G$ being the gravitational constant and $M$ the mass of the source of gravity.
The only non-vanishing tetrad components for this metric are
\be\label{stet}
e^0_{\hat{0}}=1-\phi\,, \quad e^i_{\hat{j}}=(1+\phi)\,\de^i_j\,.
\ee
Using Eqs.~(\ref{eqn:A}) and (\ref{schw}), 
one can easily prove that $A_{\mu}=0$, due
to the  spherical symmetry of the metric.
From Eq.~\eqref{eqn:pmudxmu}, it thus follows that 
the neutrino phase-shift $\varphi_{12}$ 
along a null geodesic in Schwarzschild spacetime is 
simply given by
\begin{equation}
\label{sfas}
\varphi_{12}=\frac{\Delta m^2}{2}\int^\lambda_{\lambda_0}\dr \lambda'\,,
\end{equation}
or equivalently 
\be\label{osas}
\varphi_{12}=\frac{\Delta m^2}{2}\int^\ell_{\ell_0}\frac{1+\phi}{E}\,\dr \ell'\,,
\ee
where $E\equiv P_0=g_{00}\,\dr t/\dr\lambda$ and we have employed Eq.~\eqref{dl}
to relate $\dr \lambda$ and $\dr \ell$.

Without loss of generality, we can consider the case of neutrino
radial propagation (for example, along the $x$-axis). From Eq.~\eqref{schw} we then have
\be
\dr \ell=
(1-\phi)\,\dr x\,,
\ee
so that the phase-shift $\varphi_{12}$ takes the form~\cite{Cardall}
\be\label{rp}
\varphi_{12}
=\frac{\Delta m^2}{2E}(x-x_0),
\ee
where we have exploited the fact
that $E$ is constant along the null trajectory.
Therefore, Eq.~(\ref{sp}) reads
\be\label{sps}
\Delta E\geq \frac{\mathrm{sin}^2\lf(2\theta\ri)}{T}\,\mathrm{sin}^2\lf[\frac{\Delta m^2\lf(x-x_0\ri)}{4E}\ri].
\ee
As explained above, in order to make a comparison with the corresponding flat
expression~\eqref{minkteur}, we need to rephrase
Eq.~\eqref{sps} in terms of quantities measured
by an observer at
rest in the curved background. Then, by
defining the proper distance corresponding
to the coordinate difference $x-x_0$ as
\be\label{lps}
L_p\equiv\int^x_{x_0}\hspace{-1mm}\sqrt{-g_{11}}\,\dr x'
=x-x_0+GM\ln\lf(\frac{x}{x_0}\ri),
\ee
and employing Eq.~\eqref{ene2} to express 
the asymptotic energy $E$ in terms of the local
one $E_\ell$ as
\be
E_\ell=(1-\phi)\,E,
\ee
we obtain
\be
\label{biggie}
\frac{\Delta E_\ell}{1-\phi}\geq \frac{\mathrm{sin}^2\lf(2\theta\ri)}{T}\,\mathrm{sin}^2\lf\{\frac{\Delta m^2L_p}{4 E_\ell}\left[1-\phi-\frac{GM}{L_p}\ln\left(\frac{x}{x_0}\right)\right]\ri\}.
\ee
This can be recast in a form similar to Eq.~\eqref{teurriformulation},
\be
\label{noprl}
\frac{\Delta E_\ell}{1-\phi}\geq \frac{\mathrm{sin}^2\lf(2\theta\ri)}{T}\,\mathrm{sin}^2\lf(\frac{\pi L_p}{L^{osc}}\ri)\hspace{-0.5mm},
\ee
by introducing the following definition of proper 
oscillation length $L^{osc}$~\cite{Forne}:
\be
\label{lsmald}
L^{osc}\equiv \frac{4\pi E_\ell}{\Delta m^2}\left[1+\phi+\frac{GM}{x-x_0}\ln\left(\frac{x}{x_0}\right)\right], 
\ee
where we have neglected higher-order terms in $\phi$. Moreover, we stress that the gravitational potential is to be regarded as $\phi(x)=-GM/x$.

Let us now observe that the maximization of the r.h.s. of Eq.~\eqref{noprl} yields $L_p=L^{osc}/2$, which allows us to relate the
time interval $T$ to the proper distance $L_p$
travelled by neutrinos as in Minkowski analysis. Indeed, by exploiting the condition of neutrino
null trajectory $ds^2=0$, we have 
\be\label{nts}
\dr t = \lf(1-2\phi\ri) \dr x \, , 
\ee
which leads to
\begin{equation} 
\label{nts2}
T 
= L_p\left[1+\frac{GM}{L_p}\ln\lf(\frac{x}{x_0}\ri)\right]
=  \frac{L^{osc}}{2}\lf[1+\frac{GM}{x-x_0}\mathrm{ln}\lf(\frac{x}{x_0}\ri)\ri].
\end{equation}
By inserting into Eq.~(\ref{noprl}), we finally get
\be
\label{bsc}
\Delta E_\ell
\geq  \frac{2\,\mathrm{sin}^2(2\theta)}{L^{{osc}}_{eff}(M)}\,,
\ee
where we have defined an \emph{effective oscillation length} 
$L^{{osc}}_{eff}$ depending on the mass of the source of gravity 
according to
\be \label{leffs}
L^{{osc}}_{eff}(M) \equiv L^{osc}\lf[1+\phi+\frac{GM}{x-x_0}\mathrm{ln}\lf(\frac{x}{x_0}\ri)\ri] .
\ee
with $L^{osc}$ given in Eq.~\eqref{lsmald}.
In this way, TEUR for neutrino oscillations in 
Schwarzschild spacetime acquires the same form as
in the Minkowski case, Eq.~\eqref{minkteur}. 
Note that $L^{osc}_{eff}$ is increased with respect to the corresponding expression in flat spacetime due to the gravitational field~\cite{Forne}, 
thus leading to a more stringent lower bound on the local energy uncertainty.
Clearly, for $M\rightarrow0$, 
$L^{{osc}}_{eff}(M)$ reduces to the oscillation length $L^{osc}$
 in Eq.~\eqref{oscleng}, so that
Eq.~\eqref{bsc} recovers the standard
energy condition in flat background.
We also note that, for $L^{{osc}}_{eff}(M)\to \infty$, 
we have $\Delta E_\ell\geq 0$, in accordance with the discussion
at the end of Section~\ref{NoiMs}. 

Some comments are in order here: first,
the effective oscillation length~\eqref{leffs} and the ensuing condition on neutrino
energy~\eqref{bsc} depend explicitly on the details of the metric,  as one could expect.
Such a result becomes even more interesting from the point of view 
of QFT analysis~\cite{smald}. In that context, 
indeed, neutrinos are viewed as \emph{unstable particles}, 
and Eqs.~\eqref{minkteur} and \eqref{bsc} 
are interpreted as a bound on the width of energy distribution. 
As a consequence, the quantity $L^{osc}_{eff}/(2 \sin^2(2\theta))$ 
plays the r\^{o}le of neutrino lifetime, which thus 
turns out to be intimately related to the geometric properties 
of the spacetime in which neutrinos propagate. Moreover, 
for neutrino travelling 
parallel to the $x$-axis with impact parameter $y=b$,
the condition~(\ref{leffs}) takes the form
\be\label{pms}
L^{osc}_{eff}(M) \equiv  L^{osc}\lf[1+\phi+\frac{G\hspace{0.2mm} M}{x-x_0}\mathrm{ln}\lf(\frac{r+x}{r_0+x_0}\ri)\ri],
\ee
where $\phi=-GM/r\equiv-GM/\sqrt{x^2+b^2}$. 
The above relation
will be useful for comparison with the next analysis in Lense-Thirring metric.

\subsection{Lense-Thirring metric}
The weak field solution for the gravitational field generated by a rotating spherical body on the equatorial plane is~\cite{lam}
\begin{equation}
\dr s^2=\lf(1+2\phi\ri)\dr t^2-\frac{\phi\hspace{0.2mm}\Omega}{r^2}\lf(x \, \dr y-y\, \dr x\ri) \, \dr t\\[1.5mm]\label{lt}-\lf(1-2\phi\ri)\lf(\dr x^2+\dr y^2+\dr z^2\ri) ,
\end{equation}
where $\Omega\equiv4R^2\omega/5$, $\phi$ is the gravitational potential 
defined as in Eq.~(\ref{phi}), and $\omega$ is the angular velocity of the source of gravity, which we assume to be oriented along the $z$ axis. 
This is the metric usually employed to describe
gravitomagnetic frame-dragging effect~\cite{ruggierociufolini}.

In this framework, the non-vanishing tetrad components are
\begin{equation}\label{ltet}
e^0_{\hat{0}}=1-\phi, \,\,\,\,\, e^1_{\hat{0}}=\frac{\phi\hspace{0.2mm}\Omega\hspace{0.2mm}y}{r^2}, \,\,\,\,\, e^2_{\hat{0}}=-\frac{\phi\hspace{0.2mm}\Omega x}{r^2}, \,\,\,\,\, e^i_{\hat{j}}=(1+\phi)\de^i_j.
\end{equation}
Unlike Schwarzschild case, 
$A^\mu$ has a non-trivial expression, owing to the
presence of off-diagonal contributions in the metric. Specifically, 
from Eq.~\eqref{eqn:A} it follows that
\be\label{alt}
A^\mu=\frac{\Omega}{2}\lf(0,\,-x\hspace{0.2mm}\partial_z\frac{\phi}{r^2},\,-y\hspace{0.2mm}\partial_z\frac{\phi}{r^2},\,\partial_x\frac{\phi\hspace{0.2mm} x}{r^2}+\partial_y\frac{\phi\hspace{0.2mm} y}{r^2}\ri).
\ee
Furthermore, since we are considering neutrinos 
propagating in the equatorial plane, we can set
$z=\mathrm{const}=0$ and ${\dr x^\mu_{null}}/{\dr \lambda}=({\dr x^0_{null}}/{\dr \lambda},\,{\dr x^1_{null}}/{\dr \lambda},\,{\dr x^2_{null}}/{\dr \lambda},\,0)$. Hence, 
by using Eq.~\eqref{phasep}, one can show that
the phase-shift $\varphi_{12}$ along
a null geodesic in the Lense-Thirring metric 
has the same form as in Eq.~\eqref{sfas}, since the contribution
due to the potential $A^\mu$ in Eq.~\eqref{eqn:pmudxmu} vanishes.
The remarkable difference, however, 
lies in the relation between $\dr\ell$ and $\dr\lambda$. Indeed,
by resorting to Eq.~(\ref{dl}),  we have
%
\be
\label{dlambdadl}
\dr\lambda=\dr\ell{\left[\lf(1+2\phi\ri)\hspace{-1mm}\lf(\frac{\dr t}{\dr \lambda}\ri)^2\hspace{-0.2mm}+\hspace{-0.2mm}\frac{\phi\hspace{0.2mm}\Omega \hspace{0.2mm}y}{r^2}\frac{\dr t}{\dr\lambda}\frac{\dr x}{\dr \lambda}-\frac{\phi\hspace{0.2mm}\Omega \hspace{0.2mm}x}{r^2}\frac{\dr t}{\dr\lambda}\frac{\dr y}{\dr \lambda}\right]}^{-\frac{1}{2}}\hspace{-1mm}.
\ee
Let us restrict our discussion to 
relativistic neutrinos propagating along 
the $x$-axis with impact parameter $y=b>R$. In this case, 
Eq.~\eqref{dlambdadl} reduces to
\be
\label{reduceto}
\dr\lambda=\dr\ell{\left[\lf(1+2\phi\ri)\hspace{-1mm}\lf(\frac{\dr t}{\dr \lambda}\ri)^2\hspace{-0.2mm}+\hspace{-0.2mm}\frac{\phi\hspace{0.2mm}\Omega \hspace{0.2mm}b}{r^2}\frac{\dr t}{\dr\lambda}\frac{\dr x}{\dr \lambda}\right]}^{-\frac{1}{2}}\hspace{-1mm}\,,
\ee
where $\phi=-GM/\sqrt{x^2+b^2}$.
From Eq.~\eqref{sp}, this yields 
\be\label{spsLense}
\Delta E\geq \frac{\mathrm{sin}^2\lf(2\theta\ri)}{T}\,\mathrm{sin}^2\lf[\frac{\Delta m^2\lf(x-x_0\ri)}{4E}\ri].
\ee
The above relation can be expressed in terms
of quantities measured by a local observer by employing 
the definitions~(\ref{ene2}) 
and~(\ref{lps}) of local energy and proper distance, respectively. 
In particular, we have
\begin{eqnarray}\label{lel}
E_\ell&=&\lf[1-\phi\left(1+\frac{\Omega b}{r^2}\right)\ri]E\,,\\[2mm]
L_p&=&x-x_0+GM\ln\lf(\frac{r+x}{r_0+x_0}\ri),
\end{eqnarray}
which lead to
\begin{equation}
\frac{\Delta E_\ell}{\lf[1-\phi\left(1+\Omega b/r^2\right)\ri]} \geq\frac{\mathrm{sin}^2\lf(2\theta\ri)}{T}\,\mathrm{sin}^2\lf\{\frac{\Delta m^2L_p}{4 E_\ell}\left[1-\frac{GM}{L_p}\ln\left(\frac{r+x}{r_0+x_0}\right)-\phi\left(1+\frac{\Omega b}{r^2}\right)\right]\ri\}.\label{flt}
\end{equation}
Note that Eq.~\eqref{flt} can be still cast in the form
\be
\label{flt2}
\frac{\Delta E_\ell}{\lf[1-\phi\left(1+\Omega b/r^2\right)\ri]} \geq\frac{\mathrm{sin}^2\lf(2\theta\ri)}{T}\,\mathrm{sin}^2\lf(\frac{\pi L_{p}}{L^{osc}}\ri),
\ee
provided that the oscillation length is now defined as
\be
\label{locosclen}
L^{osc}\equiv \frac{4\pi E_\ell}{\Delta m^2}\left[1+\frac{GM}{x-x_0}\ln\left(\frac{r+x}{r_0+x_0}\right)+\phi\left(1+\frac{\Omega b}{r^2}\right)\right]. 
\ee
Once again, the r.h.s. of Eq.~\eqref{flt2} is maximized 
for $L_{p}=L^{osc}/2$. Thus, 
by exploiting such a condition and performing calculations similar to the ones 
in Schwarzschild background, one can show that the time interval 
$T$ for a null trajectory is related to the 
proper distance $L_p$ travelled by neutrinos as follows:
\begin{eqnarray}\label{plt}
\non
T&=&L_p\left[1+\frac{GM}{L_p}\ln\left(\frac{r+x}{r_0+x_0}\right)+\frac{G\hspace{0.2mm}M\hspace{0.2mm}\Omega}{2\hspace{0.2mm}b\hspace{0.2mm}L_p}\lf(\frac{x}{r}-\frac{x_0}{r_0}\ri)\right]\\[2mm]
&=&\frac{L^{osc}}{2}\left[1+\frac{GM}{x-x_0}\ln\left(\frac{r+x}{r_0+x_0}\right)+\frac{G\,M\,\Omega}{2b\lf(x-x_0\ri)}\lf(\frac{x}{r}-\frac{x_0}{r_0}\ri)\right].
\end{eqnarray}
Substitution of Eq.~\eqref{plt} into Eq.~(\ref{flt2}) 
entails
\be \label{teurls}
\De E_\ell \geq \frac{2 \sin^2 (2 \theta)}{L^{osc}_{eff}(M,\Omega)} \, , 
\ee
where the effective oscillation length $L^{osc}_{eff}$ now depends on both the mass and the
angular velocity of the source according to
\begin{equation}
L^{osc}_{eff}(M,\Omega) \equiv  L^{osc}\lf[1+\frac{G \hspace{0.3mm} M}{x-x_0}\mathrm{ln}\lf(\frac{r+x}{r_0+x_0}\ri)+\frac{G\,M\,\Omega}{2b\lf(x-x_0\ri)}\lf(\frac{x}{r}-\frac{x_0}{r_0}\ri)+\phi\left(1+\frac{\Omega b}{r^2}\right)\ri] \, .
\end{equation}
Clearly, for $\Omega\rightarrow 0$, the corresponding Schwarzschild formulas~\eqref{bsc} and~\eqref{pms} are recovered, as expected for a spherically symmetric non-rotating source. 
Furthermore, we stress that, as in the previous analysis, the effective oscillation length is increased with respect to the flat case.

\subsection{Rindler metric}
Finally, let us discuss how the condition~\eqref{minkteur}
gets modified for an observer moving with eternal uniform linear acceleration.
First, we remark that, from the point of view of such an observer, 
the line element can be written
in terms of the Rindler Fermi coordinates as~\cite{misner,acc}
\be\label{acm}
\dr s^2\ =\ \lf(1+ax\ri)\dr t^2-\lf(\dr x^2+\dr y^2+\dr z^2\ri) \, ,
\ee
where we have assumed 
the acceleration $a$ to be parallel to the $x$-axis. 
Note that Eq.~\eqref{acm} only holds in the approximation
$x\ll a^{-1}$~\cite{misner,acc}.

In the above setting, one can show that
the only non-trivial tetrad component is
\be\label{ate}
e^0_{\hat{0}}=1-\frac{ax}{2},
\ee
and $A^\mu=0$. Thus, by resorting to Eqs.~\eqref{eqn:pmudxmu} and \eqref{dl}, 
the neutrino phase-shift takes the form
\be\label{ac}
\varphi_{12}=\frac{\Delta m^2}{2E}\int^x_{x_0}\lf(1+\frac{ax}{2}\ri)\dr x=\frac{\Delta m^2\,L_p}{2E_\ell}\lf(1-\frac{a}{4}L_p\ri),
\ee
where, in the second step, we have used the relations 
\begin{eqnarray}
E_\ell&=&\lf(1-\frac{ax}{2}\ri)E\,,\\[2mm]
L_p&=&x-x_0\,.
\end{eqnarray}
Note that, unlike the two previous examples,
the ``asymptotic'' energy $E$ is now to be regarded
as the energy measured by an observer at rest at the origin of the coordinate system
(where the metric Eq.~\eqref{acm} reduces to the flat Minkowski one), 
rather than the energy measured at infinity~\cite{acc}. 

From Eq.~\eqref{sp}, we have
\be\label{afp}
\frac{\Delta E_\ell}{1-ax/2} \geq \frac{\mathrm{sin}^2\lf(2\theta\ri)}{T}\,\mathrm{sin}^2\left[\frac{\Delta m^2\,L_p}{4E_\ell}\lf(1-\frac{a}{4}L_p\ri)\right]=\frac{\mathrm{sin}^2\lf(2\theta\ri)}{T}\,\mathrm{sin}^2\lf(\frac{\pi L_p}{L^{osc}}\ri)\hspace{-0.5mm},
\ee
where the oscillation length has been now defined as
\be
L^{osc} \equiv\frac{4\pi E_\ell}{\Delta m^2}\left(1+\frac{a}{4}L_p\right).
\ee
By using this equation and observing that, for null trajectories, the time interval $T$ is related to the proper distance
$L_p$ by
\be
\label{taf}
T=L_p\lf[1-\frac{a}{4}\lf(x+x_0\ri)\ri],
\ee
we finally obtain the following bound on the
local energy uncertainty:
\be
\Delta E_\ell\geq \frac{2\hspace{0.3mm}\mathrm{sin}^2(2\theta)}{L^{osc}_{eff}(a)},
\ee
where 
\be\label{losca}
L^{osc}_{eff}(a) \equiv L^{osc}\lf[1+\frac{a}{4}\lf(x-x_0\ri)\ri],
\ee
and we have imposed the usual condition $L_p=L^{osc}/2$.
Differently from Schwarzschild and Lense-Thirring geometries, however, for a given finite acceleration $a$, we cannot consider arbitrarily large 
effective oscillation lengths, since this would
violate the approximation $ax\ll 1$ in which
the metric~\eqref{acm} is valid.



\section{Concluding Remarks} \label{conc}
In this work, we have derived the time-energy uncertainty relation for neutrino oscillations in the Mandelstam-Tamm form in the case of a generic stationary background. In particular, it has been shown that the standard oscillation condition calculated in Refs.~\cite{Bil,Kayser,smald} is non-trivially modified in the presence of gravity/acceleration. By means of Stodolsky covariant definition of the quantum mechanical phase, we have evaluated the oscillation probability formula in curved spacetime, and then we have estimated lower bounds on the energy uncertainty $\Delta E$. Specifically, this has been done in 
Schwarzschild, Lense-Thirring and Rindler geometries. Deviations from the corresponding result in Minkowski spacetime have been explicitly calculated in the weak field approximation, emphasizing how the flat outcome is recovered in the limit of vanishing gravitational field/acceleration.

The aforementioned corrections to the energy uncertainty condition have been expressed by defining an effective oscillation length depending on the features of the metric itself for each treated background. 
This becomes particularly interesting if we regard such a result as the ultrarelativistic approximation of some more fundamental relation which has to be derived in the context of Quantum Field Theory. As argued in Ref.~\cite{smald}, indeed,  TEUR for neutrino oscillations in QFT can be recast in the form of a flavor-energy uncertainty principle; moreover, a field theoretical approach to the phenomenon of flavor transitions leads to a novel interpretation of the condition in Eq.~\eqref{minkteur} (and similarly for the relations obtained in curved spacetimes): flavor neutrinos can be formally viewed as \emph{unstable particles} and the quantity $(L^{osc}_{eff})/(2 \sin^2(2\theta))$ can be interpreted as the neutrino lifetime. 
The derived conditions of neutrino oscillations thus provide 
estimations of the energy distribution width, which cannot be eliminated 
in experiments and characterizes the definition of neutrino flavor states. As a result of our analysis, this definition turns out to be
closely related to the details of the spacetime in which neutrinos propagate. 
We remark that a similar achievement was
obtained in Ref.~\cite{Mueller}, where it was
shown that the decay properties of particles are less fundamental than commonly
thought.

Apart from its relevance in understanding of neutrino flavor oscillations in curved spacetime, we remark that the above  discussion
fits in the longstanding controversy about the real physical nature 
of asymptotic neutrino states in QFT~\cite{Blas95, qftmixing}.
In this sense, it should be emphasized that, whilst
in the QM treatment mixing relations between flavor
and mass neutrinos act as pure rotations on massive states,
in QFT they emerge
from the interplay between rotations and Bogoliubov 
transformations at level of ladder operators~\cite{Blas95}.
The non-commutativity between these two transformations 
turns out to play a key r\^ole, being responsible for the 
unitary inequivalence of mass and 
flavor representations and their related vacuum structures. 
Considerations based upon the 
conservation of the total lepton number in the production/detection
vertex of a weak decay process involving neutrinos~\cite{smald} seem to indicate that flavor states provide
the most appropriate description of the behavior of asymptotic neutrinos.
In that case, flavor-energy uncertainty relations 
imply that neutrino energy cannot be exactly measured, and the analysis of its 
distribution may offer valuable informations at both
theoretical and phenomenological levels.  
Similar arguments in favor of flavor states 
have been given in Refs.~\cite{protflav}
on the basis of the General Covariance of QFT
in the context of the decay of accelerated protons.

As a further development, we emphasize 
that the present analysis may open new perspectives
in the study of entanglement in curved
spacetime. Indeed, quantum information properties related to neutrino oscillations
were extensively explored~\cite{alok}. In particular, in Refs.~\cite{Blasone:2007vw,NuclPhysB} linear entropies were
employed to quantify dynamical entanglement, showing that
they are proportional to the uncertainty of the number operator
on neutrino flavor states. Such an achievement was later generalized
to QFT~\cite{qftflaventang}. It is thus evident that the study
of uncertainty relations for neutrino oscillations
is closely related to the quantum information properties of this phenomenon, as it
was already pointed out in Ref.~\cite{smald}. 
Neutrino entanglement in curved spacetime
is already under investigation.

More details about the intimate nature of neutrino mixing
and oscillations can only be understood by developing a full-fledged field theoretical treatment of these phenomena. 
Progress along this direction has been recently made in Refs.~\cite{buddell}, where 
these issues have been addressed in connection with the non-trivial structure of vacuum in QFT~\cite{nontrivial}.


\section*{Acknowledgements}
The authors would like to thank Giulio Francesco Aldi and Luca Buoninfante for helpful conversations.


\section*{References}
\begin{thebibliography}{99}
\bibitem{Pontecorvo}
  B.~Pontecorvo,
  Sov. Phys. JETP {\bf 6}, 429 (1957);
  Sov. Phys. JETP {\bf 7}, 172-02 (1957);
V.~Gribov and B.~Pontecorvo,
 Phys. Lett. B {\bf 28}, 493 (1969);
S.M.~Bilenky and B.~Pontecorvo,
Phys. Lett. B {\bf 61}, 248 (1976);
S.M.~Bilenky and B.~Pontecorvo,
Phys. Rep. {\bf 41}, 225 (1978).

\bibitem{partref}
C.~Giunti and C.~W.~Kim, \textit{Fundamentals of Neutrino Physics and Astrophysics}, (Oxford Univ. Press, Oxford, 2007);
H.~Kleinert, \textit{Particles and Quantum fields}, (World Scientific, Singapore, 2016);
A.~Strumia and F.~Vissani, 
arXiv:hep-ph/0606054v3.


 

\bibitem{Blas95}
  M.~Blasone and G.~Vitiello,
  Annals Phys.\  {\bf 244}, 283 (1995).
  \bibitem{qftmixing}
  K.~Fujii, C.~Habe and T.~Yabuki,
  Phys.\ Rev.\ D {\bf 59}, 113003 (1999);
  Phys. Rev. D {\bf 64}, 013011 (2001);
K.~C.~Hannabuss and D.~C.~Latimer,
J. Phys. A {\bf 33}, 1369 (2000);
J. Phys. A {\bf 36}, L69 (2003);
  C.~R.~Ji and Y.~Mishchenko,
  Phys.\ Rev.\ D {\bf 65}, 096015 (2002);
   C.~R.~Ji and Y.~Mishchenko,
Phys. Rev. D {\bf 65}, 096015 (2002);
  Annals Phys.\  {\bf 315}, 488 (2005);
  C.~Giunti,
  Eur.\ Phys.\ J.\ C {\bf 39}, 377 (2005);
 C.~C.~Nishi,
  Phys.\ Rev.\ D {\bf 73}, 053013 (2006);
  C. Y. Lee arXiv:1709.06306v3 [hep-ph].

  
  \bibitem{Dolgov:2002wy} 
  A.~D.~Dolgov,
  Phys.\ Rept.\  {\bf 370}, 333 (2002).
  
  
    \bibitem{Blasone:2007vw} 
  M.~Blasone, F.~Dell'Anno, S.~De Siena and F.~Illuminati,
  EPL {\bf 85}, 50002 (2009).
  
  \bibitem{Akhmedov:2010ua} 
  E.~K.~Akhmedov and A.~Y.~Smirnov,
  Found.\ Phys.\  {\bf 41}, 1279 (2011).
  
  
\bibitem{Blasone:2018hah} 
N.~E.~Mavromatos, S.~Sarkar and W.~Tarantino,
Phys. \ Rev. \ D {\bf 80}, 084046 (2009);
N.~E.~Mavromatos, S.~Sarkar and W.~Tarantino,
Mod. \ Phys. \ Lett. \ A {\bf 28}, 1350045 (2013).
M.~Blasone, P.~Jizba and L.~Smaldone,
Nuovo Cim.\ C {\bf 38}, 201 (2015);
N.~E.~Mavromatos,
PoS NEUTEL {\bf 2015}, 039 (2015);
E.~Bertuzzo, S.~Jana, P.~A.~N.~Machado and R.~Zukanovich Funchal,
arXiv:1808.02500 [hep-ph].

\bibitem{Bhat}
K.~Bhattacharyya,
J. Phys. A {\bf 16}, 2993 (1983).

 \bibitem{Bil}	
S.~M.~Bilenky,
Phys. Scripta T {\bf 127}, 8 (2006);
S.~M.~Bilenky and M.~D.~Mateev,
Phys. Part. Nucl. {\bf 38}, 117 (2007);
S.~M.~Bilenky, F.~von Feilitzsch and W.~Potzel,	
J. Phys. G {\bf 35}, 095003 (2008);
S.~M.~Bilenky, F.~von Feilitzsch and W.~Potzel,	
J. Phys. G {\bf 38}, 115002 (2011).

\bibitem{ManTam}
L.~Mandelstam and I.~G.~Tamm,
J. Phys. USSR {\bf 9}, 249 (1945).


\bibitem{Kayser}
B.~Kayser
Phys. Rev. D {\bf 24}, 110 (1981).

	\bibitem{smald}
	M.~Blasone, P.~Jizba and L.~Smaldone,
  Phys.\ Rev.\ D {\bf 99}, 016014 (2019).
	


  
  \bibitem{Stodo} 
  L.~Stodolsky,
  Gen.\ Rel.\ Grav.\  {\bf 11}, 391 (1979).
  
  
  \bibitem{Aluw} 
  D.~V.~Ahluwalia and C.~Burgard,
  Gen.\ Rel.\ Grav.\  {\bf 28}, 1161 (1996);  
  D.~Piriz, M.~Roy and J.~Wudka,
  Phys.\ Rev.\ D {\bf 54}, 1587 (1996);
  Y.~Kojima,
  Mod.\ Phys.\ Lett.\ A {\bf 11}, 2965 (1996);
  M.~Roy and J.~Wudka,
  Phys.\ Rev.\ D {\bf 56}, 2403 (1997);
  Y.~Grossman and H.~J.~Lipkin,
  Phys.\ Rev.\ D {\bf 55}, 2760 (1997);
  K.~Konno and M.~Kasai, 
  Prog.\ Theor.\ Phys.,  {\bf 100}, 1145 (1998)
  D.~V.~Ahluwalia and C.~Burgard,
  Phys.\ Rev.\ D {\bf 57}, 4724 (1998);
  T.~Bhattacharya, S.~Habib and E.~Mottola,
  Phys.\ Rev.\ D {\bf 59}, 067301 (1999);
  S.~Capozziello and G.~Lambiase,
  Mod.\ Phys.\ Lett.\ A {\bf 14}, 2193 (1999);
  M.~Adak, T.~Dereli and L.~H.~Ryder, 
   Class.\ Quant.\ Grav.\  {\bf 18}, 1503 (2001);
	G.~Lambiase,
  Class.\ Quant.\ Grav.\  {\bf 20}, 4213 (2003);
    M.~Adak, T.~Dereli and L~.H.~Ryder, 
      Phys.\ Rev.\ D {\bf 69},  123002 (2004);
      	  G.~Lambiase, G.~Papini, R.~Punzi and G.~Scarpetta,
  Phys.\ Rev.\ D, {\bf 71} 073011 (2005);
    	  G.~Lambiase,
	  Mon.\ Not.\ Roy.\ Astron.\ Soc.,  {\bf 362}, 867 (2005);
	  F.~Sorge and S.~Zilio, 
	    Class.\ Quant.\ Grav.,  {\bf 24}, 2653 (2007);
  L.~Visinelli,
  Gen.\ Rel.\ Grav.\  {\bf 47}, 62 (2015);
 S.~Chakraborty,  JCAP, {\bf 1510}, 019 (2015).

  
  
  \bibitem{Cardall}
  C.~Y.~Cardall and G.~M.~Fuller,
  Phys.\ Rev.\ D {\bf 55}, 7960 (1997).
  
    
    \bibitem{Forne}
    N.~Fornengo, C.~Giunti, C.~W.~Kim and J.~Song,
  Phys.\ Rev.\ D {\bf 56}, 1895 (1997).


\bibitem{Blasone:2018iih} 
  S.~Capozziello and G.~Lambiase,
  Eur.\ Phys.\ J.\ C {\bf 12}, 343 (2000).

  \bibitem{acc}
  M.~Blasone, G.~Lambiase, G.~G.~Luciano and L.~Petruzziello,
  EPL {\bf 124},  51001 (2018).

\bibitem{Blasone:2017nbf} 
   M.~Blasone, G.~Lambiase and G.~G.~Luciano,
  Phys.\ Rev.\ D {\bf 96}, 025023 (2017);
   M.~Blasone, G.~Lambiase and G.~G.~Luciano,
  J.\ Phys.\ Conf.\ Ser.\  {\bf 880}, 012043 (2017);
   M.~Blasone, G.~Lambiase and G.~G.~Luciano,
  J.\ Phys.\ Conf.\ Ser.\  {\bf 956},  012021 (2018).
	
\bibitem{Mueller} 
  R.~Muller,
  Phys.\ Rev.\ D {\bf 56}, 953 (1997).	

\bibitem{dw} 
  B.~S.~DeWitt,
  Phys.\ Rept.\  {\bf 19}, 295 (1975).
  
  	
\bibitem{Schumacher} 
D.~J.~Griffiths,
{\it Introduction to Quantum Mechanics}, (Prentice Hall, New Jersey, 1995);
B.~Schumacher and M.~Westmoreland, \textit{Quantum Processes, Systems, and Information} (Cambridge Univ. Press, Cambridge, 2010).

\bibitem{Akhmedov:2008zz} 
  E.~K.~Akhmedov, J.~Kopp and M.~Lindner,
  J.\ Phys.\ G {\bf 36}, 078001 (2009).
  
  
   \bibitem{gross}
   Y.~Grossman and H.~J.~Lipkin, WIS-96/27/Jun-PH, TAUP 2346-96, hep-ph/9607201
(1996).

    \bibitem{Weinberg:1995mt} 
  S.~Weinberg, \emph{The Quantum Theory of Fields}
(Cambrdige UP, Cambridge) 1995.
    

  \bibitem{lam}
	J.~Lense and H.~Thirring, Z. Phys. \textbf{19}, 156 (1918).
  
  \bibitem{ruggierociufolini} 
  I. Ciufolini and J.~A.~Wheeler, \emph{"Gravitation and Inertia"}, Princeton University Press, (1995);
  I. Ciufolini, Nature \textbf{449}, 41 (2007);
  M.~L.~Ruggiero and A.~Tartaglia,
  Nuovo Cim.\ B {\bf 117}, 743 (2002).
  
  \bibitem{misner} 
C.~W.~Misner, K.~S.~Thorne and J.~A.~Wheeler, \emph{Gravitation} (W. H. Freeman and Company, San Francisco) 1973.

\bibitem{protflav}
  M.~Blasone, G.~Lambiase, G.~G.~Luciano and L.~Petruzziello,
  Phys.\ Rev.\ D {\bf 97}, 105008 (2018);
   M.~Blasone, G.~Lambiase, G.~G.~Luciano and L.~Petruzziello,
  arXiv:1903.03382 [hep-th].

\bibitem{alok}
S.~Banerjee, A.~K.~Alok, R.~Srikanth and B.~C.~Hiesmayr,
Eur.\ Phys.\ J.\ C {\bf 75}, 487 (2015); 
A.~K.~Alok, S.~Banerjee and S.~U.~Sankar,
Nucl.\ Phys.\ B {\bf 909}, 65 (2016) .

\bibitem{NuclPhysB}
M.~Blasone, F.~Dell'Anno, S.~De Siena and F.~Illuminati,
Nucl. Phys. B (Proc. Suppl.) {\bf 237}-{\bf 238}, 320 (2013).


\bibitem{qftflaventang}
M.~Blasone, F.~Dell'Anno, S.~De Siena and F.~Illuminati,
J. \ Phys. \ Conf. \ Series {\bf 441}, 012070 (2013).

  
  \bibitem{buddell}
   M.~Blasone, F.~Dell'€™Anno, S.~De Siena and F.~Illuminati,
  Adv.\ High Energy Phys.\  {\bf 2014}, 359168 (2014);
   C.~Y.~Lee,
  arXiv:1709.06306 [hep-ph];
 M.~Blasone, G.~G.~Luciano, L.~Petruzziello and L.~Smaldone,
  Phys.\ Lett.\ B {\bf 786}, 278 (2018);
   A.~Capolupo, I.~De Martino, G.~Lambiase and A.~Stabile,
  Phys.\ Lett.\ B {\bf 790}, 427 (2019).
  
 \bibitem{nontrivial}
  V.~A.~Naumov and D.~S.~Shkirmanov,
  Mod.\ Phys.\ Lett.\ A {\bf 30}, 1550110 (2015);
  M.~Blasone, G.~Lambiase, L.~Petruzziello and A.~Stabile,
  Eur.\ Phys.\ J.\ C {\bf 78}, 976 (2018);
  M.~Blasone, P.~Jizba and G.~G.~Luciano,
  Annals Phys.\  {\bf 397}, 213 (2018);
  F.~Scardigli, M.~Blasone, G.~G.~Luciano and R.~Casadio,
  Eur.\ Phys.\ J.\ C {\bf 78}, 728 (2018);
   L.~Buoninfante, G.~Lambiase, L.~Petruzziello and A.~Stabile,
  Eur.\ Phys.\ J.\ C {\bf 79}, 41 (2019);
 G.~G.~Luciano and L.~Petruzziello,
  Eur.\ Phys.\ J.\ C {\bf 79}, 283 (2019).
 
	\end{thebibliography}
\end{document}